# Feasibility of dual-energy CBCT material decomposition in the human torso with 2D anti-scatter grids and grid-based scatter sampling

Cem Altunbas

Department of Radiation Oncology, University of Colorado School of Medicine, Aurora, CO

**Email:** cem.altunbas@cuanschutz.edu

**Address:** 1665 Aurora Court, Suite 1032, Aurora, CO, 80045

**Background:** Dual-energy (DE) imaging techniques in cone-beam computed tomography (CBCT) have potential clinical applications, including material quantification and improved tissue visualization. However, the performance of DE CBCT is limited by the effects of scattered radiation, which restricts its use to small object imaging.

**Purpose:** This study investigates the feasibility of DE CBCT material decomposition by reducing scatter with a 2D anti-scatter grid and a measurement-based scatter correction method. Specifically, the investigation focuses on iodine quantification accuracy and virtual monoenergetic (VME) imaging in phantoms that mimic head, thorax, abdomen, and pelvis anatomies.

**Methods:** A 2D anti-scatter grid prototype was utilized with a residual scatter correction method in a linac-mounted CBCT system to investigate the effects of robust scatter suppression in DE CBCT. Scans were acquired at 90 and 140 kVp using phantoms that mimic head, thorax, and abdomen/pelvis anatomies. Iodine vials with varying concentrations were placed in each phantom, and CBCT images were decomposed into iodine and water basis material images. The effect of a 2D anti-scatter grid with and without residual scatter correction on iodine concentration quantification and contrast visualization in VME images was evaluated. To benchmark iodine concentration quantification accuracy, a similar set of experiments and DE processing were also performed with a conventional multidetector CT scanner.

**Results:** In CBCT images, a 2D grid with or without scatter correction can differentiate iodine and water after DE processing in human torso-sized phantom images. However, iodine quantification errors were up to 10 mg/ml in pelvis phantoms when only the 2D grid was used. Adding scatter correction to 2D-grid CBCT reduced iodine quantification errors below 1.5 mg/ml in pelvis phantoms, comparable to iodine quantification errors in multidetector CT. While a noticeable contrast-to-noise ratio improvement was not observed in VME CBCT images, contrast visualization was substantially better in 40 keV VME images in visual comparisons with 90 and 140 kVp CBCT images across all phantom sizes investigated.

**Conclusions:** This study indicates that accurate DE decomposition is potentially feasible in DE CBCT of the human torso if robust scatter suppression is achieved with 2D anti-scatter grids and residual scatter correction. This approach can potentially enable better contrast visualization and tissue and contrast agent quantification in various CBCT applications.

# 1. Introduction

DE imaging, or spectral imaging in general, is one of the major techniques to increase the information content of multi-detector CT (MDCT) images.[1] DE material decomposition into 2 or 3 basis materials allows the differentiation of tissues and contrast agents with different elemental compositions. Low contrast visualization can also be improved by synthesizing virtual monoenergetic images in DE MDCT.[2,3] A natural progression in DE imaging is the translation of DE techniques from MDCT to CBCT systems.[4] DE CBCT can potentially bring various benefits in different clinical applications. For instance, in breast CBCT, DE techniques can enhance the measurement of breast composition.[5] In transcatheter arterial chemoembolization (TACE) procedures, DE CBCT can provide better visualization and localization of tumors by improving the enhancement of iodine.[6] DE techniques have been explored in extremity imaging to enhance the detection of subtle fractures and bone marrow edema.[7] DE imaging in CBCT-guided radiation therapy can potentially improve the visualization of low-contrast targets, enhance the accuracy of dose calculations during adaptive photon or proton therapy, reduce metal artifacts, and increase the precision of radiomics features that predict tumor control and toxicity.[8-11] Additionally, DE techniques may allow for novel approaches in CBCT-guided therapy, such as contrast-enhanced CBCT imaging, which could further enhance the visualization of soft tissue targets.[12]

While DE techniques have been increasingly used in MDCT systems for more than a decade,[1] DE imaging in clinical flat panel detector-based CBCT systems has not been available yet due to two major challenges. The first challenge is the acquisition of high and low energy projections at high temporal resolution. Due to slow gantry rotation in CBCT systems and the relatively slow frame acquisition rate of flat panel detectors, high and low-energy projections cannot be acquired in a short time frame to eliminate motion-induced inconsistencies between low and high-energy scans. Therefore, most of the DE CBCT research has been focused on developing new DE acquisition techniques to address this challenge. Numerous approaches have been proposed to acquire projections at multiple energy ranges. For example, fast kV switching during CBCT scans can allow rapid acquisition low and high kV projections during one CBCT scan.[11,13] Another approach is to utilize a constant kV CBCT scan and employ pre-patient beam filters to modulate the beam energy incident on the patient.[14] Dual or multilayer flat panel detectors were also proposed recently, where a constant kV is used during a CBCT scan, and low-high energy X-rays are detected by different layers of the detector.[15-17] Finally, photon counting detectors are promising in their potential to differentiate the energy of individual photons using one detector in a constant kV scan.[18-20]

The second major challenge in DE CBCT involves addressing the poor accuracy of X-ray transmission measurements, which can degrade the fidelity of the DE processing. Scattered radiation is one of the primary causes of this challenge, resulting from the inherent problem of large cone angles in flat-panel detector-based CBCT. To perform DE processing accurately, it is necessary to obtain X-ray transmission measurements that are quantitatively precise. However, measurement accuracy is often compromised due to signal bias and noise introduced by scattered X-rays.[21-23] To minimize the detrimental effects of scatter during DE processing, current research and applications in DE CBCT have primarily focused on imaging small objects and extremities,[4,7,11,15,16] or on using lower cone angles for the X-ray beam.[18,20]

Therefore, the application of DE CBCT imaging in the human torso remains a major challenge due to the scattered radiation problem. Robust solutions for scatter mitigation are needed to enable DE CBCT imaging of the human torso. Several methods have been investigated, including radiographic anti-scatter grids,[16,24] scatter correction methods,[11] deep learning methods,[25,26] and primary modulation methods.[27-29]

This study proposes and investigates a new approach for suppressing scatter in DE CBCT imaging of the human torso. The approach combines a two-dimensional anti-scatter grid, currently under development, with a measurement-based scatter correction method known as Grid-based Scatter Sampling (GSS). Previous studies have shown that both 2D anti-scatter grids and GSS can improve quantitative accuracy in conventional, single-energy CBCT imaging for image-guided radiation therapy applications.[30-34] To investigate the effect of scatter suppression on iodine concentration quantification accuracy and contrast enhancement in VME images, DE CBCT images of head, thorax, and abdomen/pelvis phantoms were evaluated.

## 2. Materials and Methods

### 2.1. Scatter suppression with 2D anti-scatter grid and grid-based scatter sampling

Although a 2D anti-scatter grid can reject the majority of scattered X-rays, approximately 5-10% of scatter fluence can still pass through the grid, resulting in a degradation of CT number accuracy in CBCT images.[34-36] Therefore, a combined scatter rejection and correction approach was explored to improve DE processing. To reject scatter, we used a focused 2D anti-scatter grid prototype with a grid pitch of 2 mm, grid ratio of 12, and wall thickness of 0.1 mm. The prototype was fabricated from pure tungsten using a selective laser melting process and was installed directly on the flat panel detector in the linac-mounted CBCT system.[36]

To correct residual scatter transmitted through the 2D grid, a GSS method was employed, which was explained in detail in prior publications.[32,33] In brief, the GSS approach uses the 2D grid as a scatter measurement device, with the grid wall shadows acting as micro fluence modulators. When scatter is present, the ratio of image signals in the grid shadows and grid holes changes depending on the intensity of residual scatter. This is because scatter is an additive signal, while reduced primary signal in grid shadows is a multiplicative effect. The GSS method takes advantage of this signal difference to estimate and correct scatter.

To explain the GSS formalism, a phantom projection with a 2D grid in place and no residual scatter present should be considered first. In a small neighborhood of pixels (e.g., 2×2 mm$^2$), the primary image signal residing in grid holes, $P_{hole}(u,v)$ are larger than the signal in septal shadows, $P_{wall}(u',v')$, because pixels residing in septal shadows receive fewer primary x-rays due to radiopaque grid septa. Assuming that primary signal variations due to phantoms can be omitted in a small neighborhood, primary signal difference between grid holes and adjacent grid wall shadows can be compensated, or equalized, by using a Gain Map, *GM,* correction.

$$P_{hole}(u,v)\, GM_{hole}(u,v) = P_{wall}(u',v')\, GM_{wall}(u',v') \qquad (1)$$

For any arbitrary pixel (*i,j*) in a projection, *GM* is

$$GM(i,j) = \frac{C}{F(i,j)} \qquad (2)$$

*F* is a flood projection acquired without an object but with a 2D grid on the flat panel detector, and *C* is an arbitrary normalization constant. In this work, *C* was the average of pixel values in the flood projection. In essence, *GM* characterizes signal variations introduced by the 2D grid's footprint, or wall shadows, in flood projections.

When residual scatter is present in a projection, it appears as a slowly varying and additive signal. Since GM correction is a multiplicative correction and scatter signal is additive, GM correction would not equalize signals residing in grid holes and adjacent grid shadows. After GM correction, signal difference *d* at a wall shadow location with respect to the adjacent grid hole would be,

$$d(u',v') = \left[(P(u',v') + S(u',v'))_{wall}\right] GM(u',v')_{wall} - \left[(P(u,v) + S(u,v))_{hole}\right] GM(u,v)_{hole} \tag{3}$$

Assuming that scatter signal intensity is uniform in the grid wall shadow and adjacent grid holes, scatter signal *S* at the grid wall location can be calculated[33] from Eqs. 1 and 3

$$S(u',v') = \frac{d(u',v')}{GM_{wall}(u',v') - \langle GM_{hole}(u,v)\rangle} \tag{4}$$

$\langle GM_{hole}\rangle$ was obtained by averaging *GM* values corresponding to grid holes in the small pixel neighborhood surrounding the pixel of interest residing in a wall shadow.

Thus, the signal difference, *d*, was calculated between each pixel residing in wall shadows and neighboring grid holes in a projection. Since each pixel in a wall shadow was surrounded by multiple pixels residing in grid holes, *d* was calculated with respect to all grid hole pixels in a small neighborhood (2×2 mm$^2$) and averaged. This process was repeated for all pixels in the wall shadows. Subsequently, the value of *S* was calculated for pixels residing in the grid shadows, and scatter for pixels in the grid holes was calculated via 2D interpolation. An interpolated 2D scatter map was obtained for each pixel (*i,j*), and this map was then subtracted from the phantom projection to correct scatter and obtain the primary-only projection.

$$P(i,j) = T(i,j) - S(i,j) \tag{5}$$

Where *T(i,j)* is the raw phantom projection (primary + scatter), scatter correction was followed by GM correction to suppress the grid shadows.

$$P_{GMCorr}(i,j) = P(i,j) \times GM(i,j) \tag{6}$$

This process was repeated for each projection. In addition to correcting residual scatter, residual scatter-to-primary ratio (SPR) in 2D grid CBCT projections was also calculated using this method.

### 2.2. Dual-energy processing

DE processing was performed in the image domain by acquiring two CBCT or MDCT scans at 90 and 140 kVp, where each voxel was expressed as a linear combination of two basis materials.

$$\begin{bmatrix} \mu_{1,Low} & \mu_{2,Low} \\ \mu_{1,High} & \mu_{2,High} \end{bmatrix} \begin{bmatrix} C_1 \\ C_2 \end{bmatrix} = \begin{bmatrix} \mu_{Low} \\ \mu_{High} \end{bmatrix} \tag{7}$$

$\mu_{Low}$ and $\mu_{High}$ are linear attenuation coefficients in a voxel reconstructed from low and high kVp scans, respectively. Two basis materials have effective mass attenuation coefficients $\mu_1$ and $\mu_2$ at low/high kVp, and concentrations $C_1$ and $C_2$, respectively. Mass attenuation coefficients of basis materials are measured in low/high kVp calibration phantom scans, where known concentrations of basis materials are present. Unknown concentrations of basis materials in a DE scan are extracted via matrix inversion,

$$\begin{bmatrix} C_1 \\ C_2 \end{bmatrix} = \begin{bmatrix} \mu_{1,Low} & \mu_{2,Low} \\ \mu_{1,High} & \mu_{2,High} \end{bmatrix}^{-1} \begin{bmatrix} \mu_{Low} \\ \mu_{High} \end{bmatrix} \tag{8}$$

In this work, iodine and water were selected as the basis materials. Mass attenuation coefficients of iodine and water were obtained by CT scanning iodine and water vials with known concentrations placed in a 20 cm diameter cylindrical phantom. Subsequently, the DE image was decomposed into iodine and water images from voxel-specific concentrations of iodine and water calculated from Eq. 8.

To obtain virtual monoenergetic images (VME) at energy *E*, water and iodine images obtained via Eq. (8) were weighted with energy-specific mass attenuation coefficients and summed.[3]

$$VME(E) = C_1 \frac{\mu}{\rho}(E)_1 + C_2 \frac{\mu}{\rho}(E)_2 \tag{9}$$

VME images were used to selectively enhance the visualization of iodine in the CT images. Another approach to differentiate materials with a given composition in DE images involves analyzing the correlation of attenuation coefficients, or CT numbers, in low and high kVp images. For instance, the ratios of attenuation coefficients in low/high kVp CBCT images are unique for bone and iodine contrast agents due to their differences in elemental composition. To calculate an attenuation coefficient ratio unique to iodine, iodine-containing vials were scanned at 90 and 140 kVp. This ratio was then used to filter iodine-containing voxels more accurately in the DE images of the test phantoms.

DE processing was performed on both 2D grid-only and 2D grid+GSS CBCT images to investigate the effect of both scatter suppression approaches on the DE CBCT performance.

**2.3. Experiment setup**

Experiments were performed using a Varian TrueBeam CBCT (Varian Medical Systems, Palo Alto, CA), a linac-mounted CBCT system for image-guided radiation therapy. For each phantom under investigation, two sets of CBCT acquisitions were sequentially performed at 90 and 140 kVp. A 0.89 mm titanium filter was in place during both acquisitions. A bow tie filter was not employed to avoid beam hardening effects caused by the aluminum bow tie filter. The CBCT system was operated in offset detector geometry to increase the field of view, and a total of 900 projections were acquired per scan. Source-to-detector and source-to-isocenter distances were 150 and 100 cm, respectively. The radiation exposure field of view was 33 × 43 cm$^2$ at the detector plane, emulating the clinical imaging and scatter conditions in CBCT-guided radiation therapy.

Two subsets of data were generated from each DE CBCT scan, corresponding to two different scatter suppression strategies: 1) scatter suppression with the 2D grid only, 2) scatter suppression with the 2D grid and correction of remaining scatter with the GSS method. All datasets were flat field corrected, and a water equivalent beam hardening correction was applied.

As a reference, a subset of phantoms was scanned with an MDCT for radiation therapy treatment planning (Philips Big Bore 16 slice CT scanner, Netherlands), where 90 and 140 kVp scans were acquired sequentially. CTDIvol values in MDCT scans were 11.6 mGy and 22.9 mGy for 90 and 140 kVp scans, respectively. CTDIvol values of CBCT scans were 7 and 9 mGy for 90 and 140 kVp scans, respectively. CBCT imaging doses were kept relatively low to prevent detector saturation at the phantom-air boundary.

CBCT projections were acquired at 0.388×0.388 mm$^2$ pixel size by using the dynamic gain mode of the flat panel detector.[37] Projections were 3×3 binned, and images were reconstructed using the FDK method modified for offset detector geometry.[38,39] Voxel size was 0.9×0.9 mm$^2$ in the axial plane, and the slice thickness was 1 mm. MDCT images were reconstructed using filtered backprojection, and the voxel size was 0.9×0.9 mm$^2$ in the axial plane, and the slice thickness was 3 mm. To achieve comparable noise levels in CBCT and MDCT images, 10 CBCT slices were averaged after DE processing and used in analyses.

A 20 cm diameter polymethyl methacrylate (PMMA) cylinder was fabricated and employed as the calibration phantom for DE processing. Iodine vials with a diameter of 2.5 cm and concentrations of 2, 5, 10, 15, 20, 30, and 40 mg/ml were prepared by diluting Isovue 370 Iopamidol (Bracco Diagnostics, Italy).

Five phantoms were employed to study the effect of object size and composition-dependent scatter fluence on the DE processing performance (Table 1). These phantoms mimicked head, thorax, and abdomen/pelvis anatomies.

To evaluate the effect of each phantom on scatter fraction, scatter was measured in each projection using the GSS method, and SPR was calculated. Average SPR was measured in a 1×1 cm$^2$ region of interest placed at the piercing point in each projection.

**Table I.** A summary of phantoms and their properties used in DE experiments

| Phantom name | Body material properties | Cross section | Dimensions in axial plane | Length in axial direction |
|---|---|---|---|---|
| Head H$_2$O | Water equivalent | Cylinder | 20 cm in diameter | 16 cm |
| Head PMMA (calibration) | PMMA | Cylinder | 20 cm in diameter | 20 cm |
| Thorax | Water and lung equivalent | Thorax | 20 × 30 cm$^2$ | 30 cm |
| Pelvis PMMA | PMMA and water equivalent | Ellipse | 30 × 38 cm$^2$ | 20 cm |
| Pelvis H$_2$O | Water equivalent | Ellipse | 30 × 40 cm$^2$ | 16 cm |

## 2.4 Data analysis

After DE material decomposition, iodine concentration measurement accuracy was evaluated in CBCT and MDCT images. Iodine concentrations were measured by placing circular ROIs in iodine-containing vials, with the ROI diameter set at 60% of the diameter of each iodine vial.

The impact of VME imaging on CNR, contrast, and noise was assessed by employing the 15 mg/ml iodine vial in the head and pelvis H$_2$O phantoms. To measure contrast, two regions of interest (ROIs) were positioned, one within the vial and the other adjacent to the vial in the water equivalent background.

$$CNR = \frac{HU_{contrast} - HU_{background}}{(\sigma_{contrast} + \sigma_{background})/2} \tag{10}$$

where contrast is the numerator and noise is the denominator of the CNR expression. $HU_{contrast}$ and $HU_{background}$ are the mean HU values in the iodine vial and the background, respectively. $\sigma_{contrast}$ and $\sigma_{background}$ are the standard deviations of HU values, or noise, in the iodine vial and the background.

In addition to CNR, peak signal-to-noise ratio (PSNR) was calculated in the VME images of head and pelvis H$_2$O phantoms.

$$PSNR = 10 \times log_{10}\left(\frac{I_{max}^2}{MSE}\right) \tag{11}$$

where $I_{max}$ is the maximum pixel intensity in the iodine vial in the image of interest, and MSE is the mean squared error in the iodine vial with respect to a reference image, Ref,

$$MSE = \frac{1}{mn}\sum_{i=0}^{m-1}\sum_{j=0}^{n-1}[I(i,j) - Ref(i,j)]^2 \tag{12}$$

In this study, Ref was the mean of 90 and 140 kVp CBCT images. Subsequently, the relative change in PSNR was calculated with respect to the minimum PSNR for each phantom and scatter suppression configuration.

$$Relative\ PSNR = PSNR - PSNR_{min} \qquad (13)$$

## 3. Results

In the CBCT projections obtained using the 2D grid, the SPR values for the head and thorax phantoms were all below 0.2 in all projections (Fig. 1). However, for the ellipse-shaped pelvis phantoms, the SPR varied between 0.4 and 1.6 depending on the gantry angle. The average SPR for the pelvis PMMA and $H_2O$ phantoms was 0.45 and 0.8, respectively. The differences in SPR values between the 90 and 140 kVp scans were minimal, which is consistent with previous SPR measurements made using 2D grids.[35]

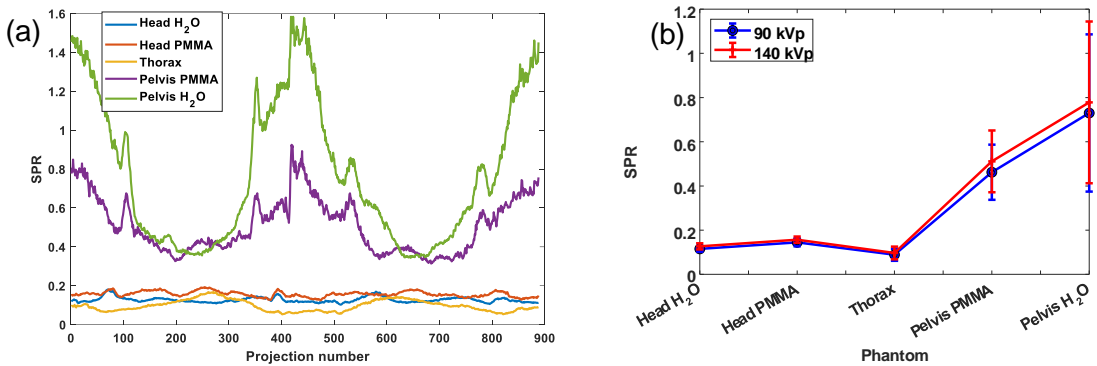

Fig. 1. (a) SPR in a 1×1 cm² ROI placed at the piercing point in projections as measured by the GSS method. (b), Mean (averaged over all projections) and standard deviation of SPRs in 90 and 140 kVp scans.

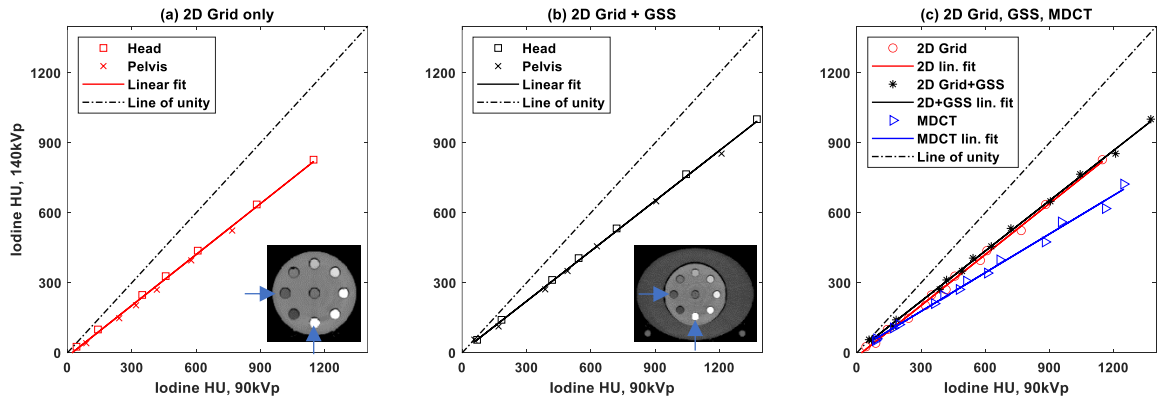

Fig. 2. HU correlation of iodine vials at 90 and 140 kVp in head and pelvis PMMA phantoms are shown for (a) 2D grid only and (b) 2D grid+GSS methods. (c) shows all methods including the MDCT. Linear fits for each imaging method were generated by combining data from head and pelvis PMMA phantoms. Iodine concentrations ranging from 2 to 40 mg/ml were used for the fits. Vials from lowest to highest concentrations were placed in the head phantom, from 9 to 6 o'clock positions in the clockwise direction, as shown by the inset in (a). For the pelvis phantom, the same vial arrangement was used, but with the addition of annulus, as shown by the inset in (b).

Iodine HU correlations between 90 and 140 kVp scans were linear for both MDCT and CBCT images (Fig. 2). Even though 2D grid-only CBCT images had relatively high SPR values for the pelvis phantom, it did not cause a deviation in the linearity of the HU correlation. This was attributed to the similarity of SPR values in the 90 and 140 kVp scans (Fig. 1), which caused a linearly proportional reduction in HU values in low and high kVp scans. HU correlation of MDCT scans was the furthest from the line of unity, implying that energy separation between 90 and 140 kVp scan was the highest in MDCT images.

In 2D grid+GSS images, iodine HU values in the head phantom were higher than in the pelvis phantom by 2.5% and 7.4% in 5 and 40 mg/ml iodine vials, respectively. In 2D grid-only images, HU values in the head phantom images were higher by 5.4% and 16.6% in 5 and 40 mg/ml iodine vials, respectively.

The linear fits applied to the iodine data produced an attenuation coefficient ratio specific to iodine for the 90/140 kVp scans. This ratio was then utilized to distinguish iodine from other high-density materials in the CBCT images, as elaborated below.

### 3.1. Iodine and water decomposition in DE images

In both CBCT and MDCT images of the head and pelvis PMMA phantoms, water and iodine were clearly distinguishable (Fig. 3). As the head PMMA phantom was used for DE calibration, a relatively accurate separation of water and iodine was expected. In the pelvis PMMA phantom, despite high SPR due to scatter in projections, water and iodine were still well-differentiated.

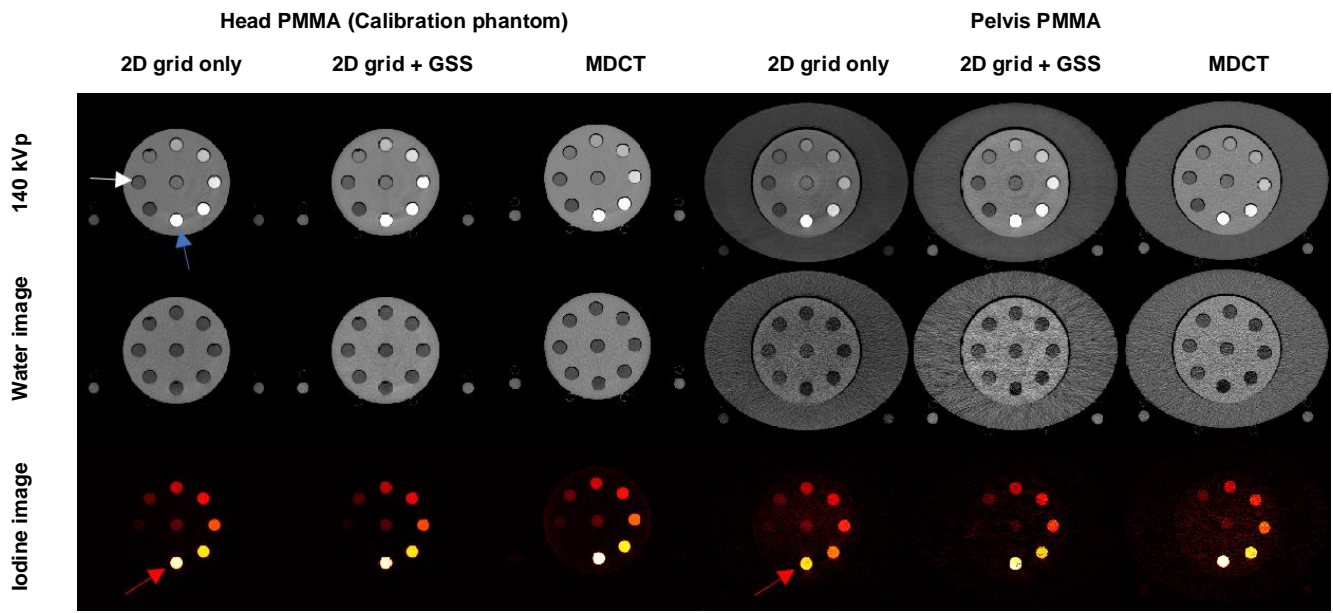

Fig. 3. Water and iodine decomposition using two different phantom sizes. Head phantom was used as the calibration phantom to extract energy decomposition parameters. Iodine vials have concentrations of 2, 5, 10, 15, 20, 30, and 40 mg/ml (from white to blue arrow in clockwise direction). Central vial has 5 mg/ml iodine concentration. Iodine concentration in the 40 mg/ml vial appears lower (red arrows), when phantom size is increased from head to pelvis. 140 kVp window: [-200 500] HU, Water window: [800 1500] mg/ml, Iodine window: [0 40] mg/ml

However, there were substantial differences in measured iodine concentrations between the head and pelvis phantom 2D grid-only CBCT images (as shown in Fig. 4). Specifically, the iodine concentration in the pelvis phantom was found to be 9.6 mg/ml less than the value for the same vial measured in the head phantom (for a 40 mg/ml vial). Furthermore, the error in measured iodine concentrations increased monotonically as a function of the nominal iodine concentration. However, these iodine quantification errors were mitigated when the 2D grid was used in conjunction with the GSS residual scatter correction method. The maximum difference in iodine concentration between the head and pelvis phantoms was only 1.5 mg/ml for the 2D grid+GSS and 0.9 mg/ml for the MDCT images, respectively. Ideally, the difference between the head and pelvis phantoms should be zero, as measured iodine concentration should not depend on the phantom size.

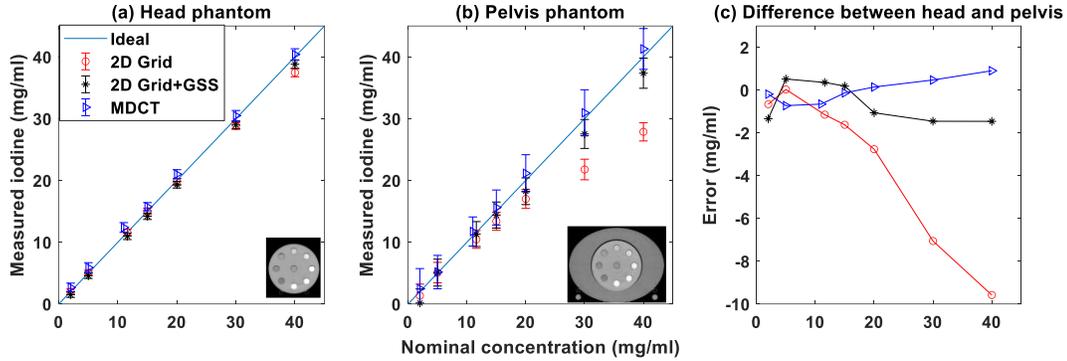

Fig. 4. Measured iodine versus nominal concentrations in (a) head PMMA and (b) pelvis PMMA phantoms. Errors in measured iodine concentrations due to changes in phantom size are shown in (c).

Due to the lower concentration of iodine (5-20 mg/ml), the water concentration remained close to its nominal value (1000 mg/ml) in iodine vials, as evidenced by the water concentration maps of the head phantom. However, in the pelvis phantom, the water concentration was underestimated, particularly for vials containing larger concentrations of iodine, due to the underestimation of HU values in the 90 and 140 kVp pelvis phantom images. In 15 mg/ml iodine vials, the mean water concentrations were lower than the nominal value by 8.7% and 4.1% in 2D grid-only and 2D grid+GSS images of the pelvis phantom, respectively.

Noise amplification in the water image was noticeable for 2D grid+GSS, which was due to higher noise after scatter correction. Similar results were obtained in both head and pelvis phantoms for both CBCT and MDCT images.

To further explore anatomy-dependent DE imaging performance, water and iodine decomposition was performed in head $H_2O$, thorax, and pelvis $H_2O$ phantom images. 2, 5, 10, and 15 mg/ml iodine vials were placed in these phantoms (Fig. 5), and iodine concentration in 15 mg/ml vial was measured (Fig. 6). Average SPR was below 0.15 in head $H_2O$ and thorax phantoms, and as a result, iodine was differentiated well. Measured iodine concentrations were within 1mg/ml of nominal values in both 2D grid and 2D grid+GSS images.

The average SPR reached 0.77 in the pelvis $H_2O$ phantom projections, leading to an error of 2.5 mg/ml in measured iodine concentration in the 2D grid-only CBCT images. However, after scatter correction with the GSS method, the error was reduced to below 1 mg/ml. Additionally, iodine-water decomposition was qualitatively less accurate in the 2D grid-only image. The phantom appeared to have low levels of iodine in the background, even though no iodine was present, resulting in iodine quantification errors due to inaccuracies in material decomposition.

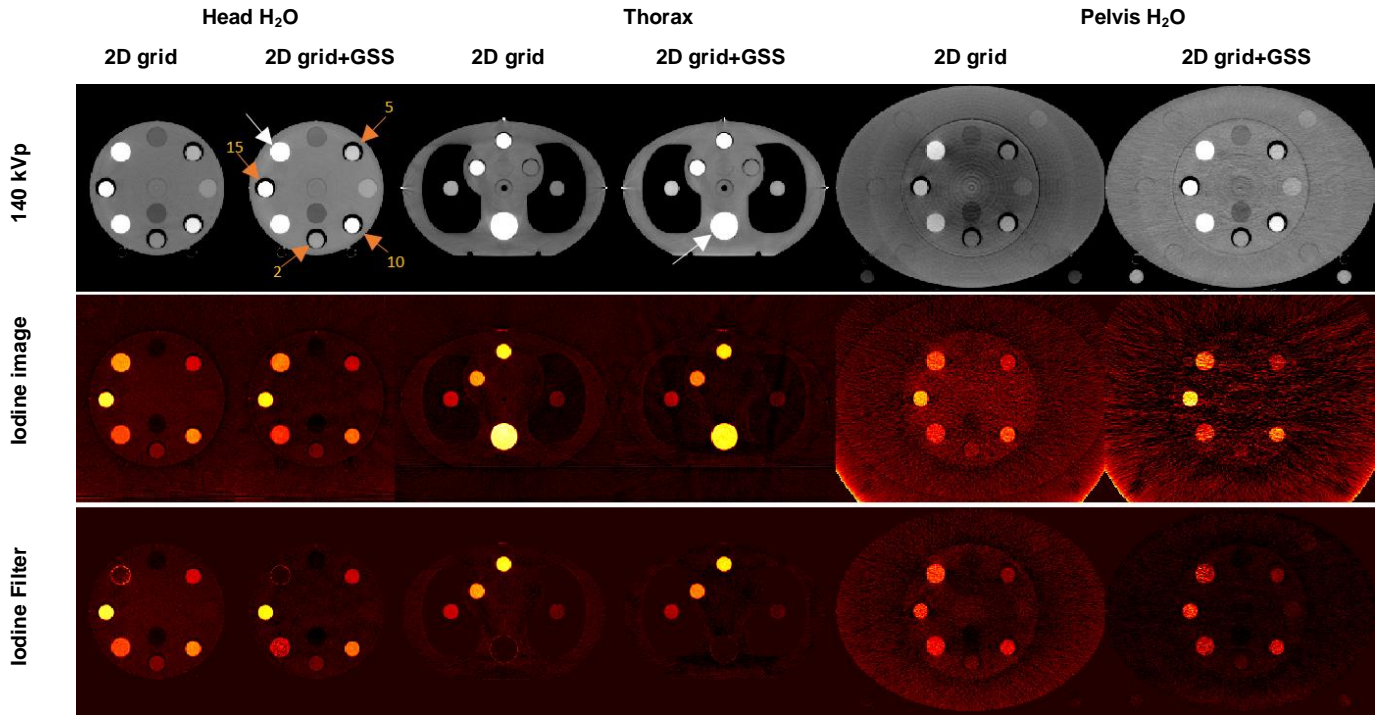

Fig. 5. Iodine images after DE decomposition in the head, thorax, and large pelvis phantom CBCT images. Iodine vials with 2, 5, 10, and 15 mg/ml concentrations were placed in each phantom (orange arrows). Additional filtering based on HU correlation maps eliminates some of the bone-like objects from iodine images (white arrows). HU window: [-250 250] HU, Iodine window: [-1 20] mg/ml.

An alternative approach to differentiate voxels containing iodine is to use the correlation between iodine HU values in 90 and 140 kVp scans, as shown in Fig. 7. By selecting voxels within the 90% confidence intervals in the HU correlation plot, erroneous iodine background signal was reduced in the iodine images (Fig. 5). Additionally, this filtering method removed some bone-like objects from the iodine images (white arrows in Fig. 5). This was partly due to the better separation between iodine and non-iodine clusters in the HU correlation plots. For instance, in the 2D grid-only pelvis $H_2O$ phantom plot (Fig. 7(c)), bone clusters are mixed with iodine clusters. However, in the 2D grid+GSS HU correlation plot (orange arrows, Fig. 7(f)), iodine voxel clusters are more visible and separated from bone voxel clusters (blue arrows, Fig. 7(f)).

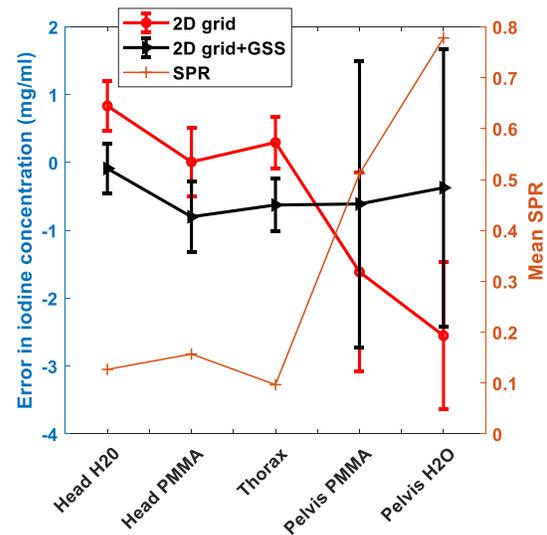

Fig. 6. Difference between measured and nominal iodine concentrations for 15 mg/ml iodine vial across all phantoms investigated. The mean SPR (average of SPRs over all projections as shown in Fig.1(b)) was overlaid to indicate the correlation between errors in measured iodine concentrations and the mean SPR.

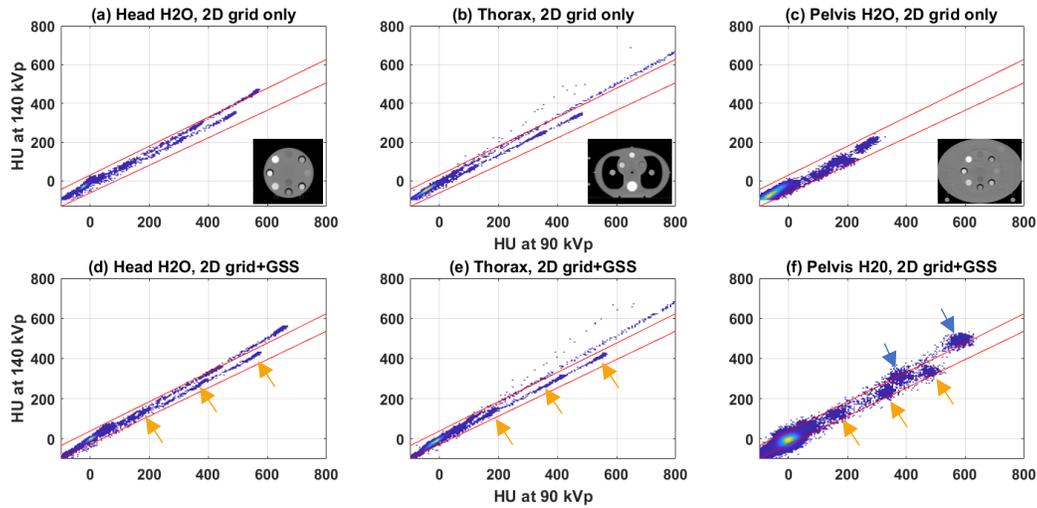

Fig. 7. HU correlation between 90 and 140 kVp CBCT scans for head $H_2O$, thorax, and pelvis $H_2O$ phantoms. Red lines indicate 90% confidence intervals obtained from linear fits to iodine vial HU values in PMMA phantoms (Fig. 2), indicating the region likely to be the iodine basis material. 5, 10, and 15 mg/ml iodine voxels are identified in the 2D grid+GSS CBCT images (orange arrows). Likewise, voxels belonging to two bony objects are visible (blue arrows).

### 3.2. Contrast visualization in VME images

VME image synthesis was explored to selectively improve iodine visualization (see Fig. 8). The CNR enhancement trends in the 2D grid-only and 2D grid+GSS VME images were comparable (see Fig. 9(a)). The CNR improvement was similar for both small and large phantoms. In general, the CNR was maximized around 50 keV VME images for all phantoms, which is above the K edge of iodine. However, the CNR at 50 keV was not higher than the CNR at 90 kVp. This issue was primarily caused by noise amplification in VME images at or above the K edge of iodine. While the contrast was improved by a factor of 3 at 40 keV (see Fig. 9(b)), the noise was amplified by the same amount or more (see Fig. 9(c)), preventing CNR increase in VME images. Even though the maximum contrast increase was at 40 keV, the maximum CNR was achieved at 50–60 keV due to lower noise at such energies. The highest CNR was achieved in either 90 kVp or averaged 90+140 kVp images among polyenergetic CBCT images.

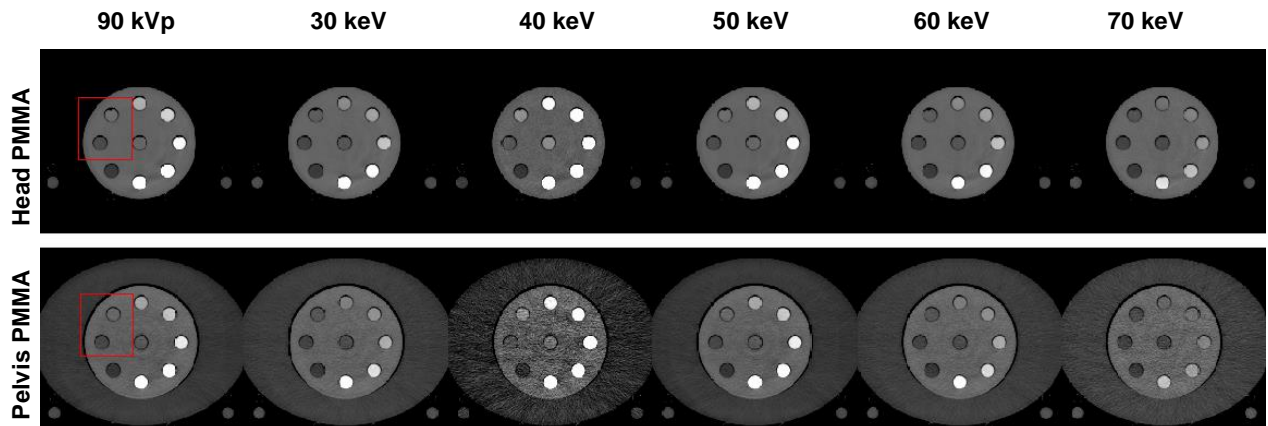

Fig. 8. VME images of head and pelvis PMMA phantoms with iodine vials. 2 and 5 mg/ml vials in red rectangles are shown in Fig. 10. HU window [-200 700].

Despite lower CNRs in 40 keV VME images than polyenergetic images, iodine visualization was qualitatively better in 40 keV VME images due to higher iodine contrast in VME images. For example, 2 and 5 mg/ml iodine vials were visualized better in 40 keV head and pelvis PMMA phantom images (Fig. 10). 5 mg/ml iodine vial (red arrow) has an appearance similar to the PMMA background in 90 and 140 kVp images, whereas the iodine vial is clearly visible in the VME image. Visualization of iodine contrast was comparable in the 2D grid and 2D grid+GSS VME images. Similar iodine contrast enhancement trends were observed in the thorax, head $H_2O$ and pelvis $H_2O$ phantoms (Fig. 11).

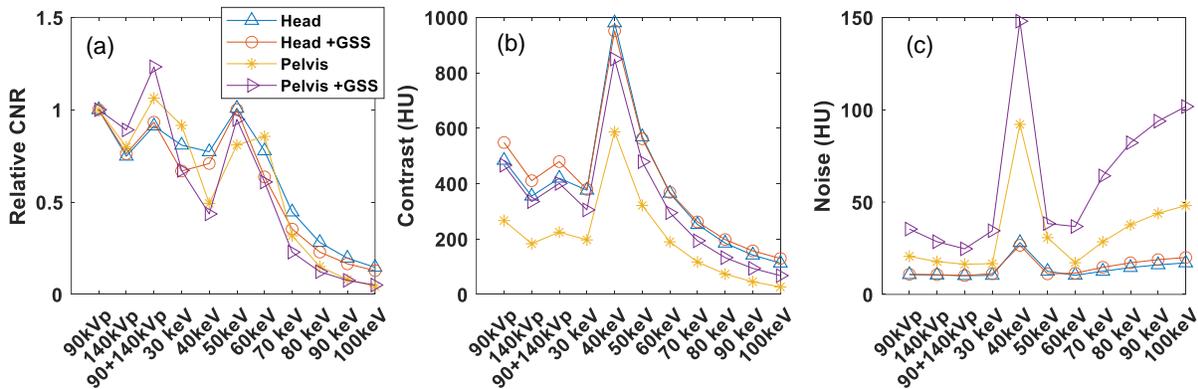

Fig. 9. Relative CNR, contrast, and noise for CBCT images of head and pelvis $H_2O$ phantoms. CNRs are normalized to CNR at 90 kVp for each respective image data set. All values were measured using 15 mg/ml iodine vial.

The relative improvement in PSNR followed the trends in CNR (Fig. 12). PSNR was minimized at 40 keV due to high noise, and it was maximized in the range of 50 to 60 keV.

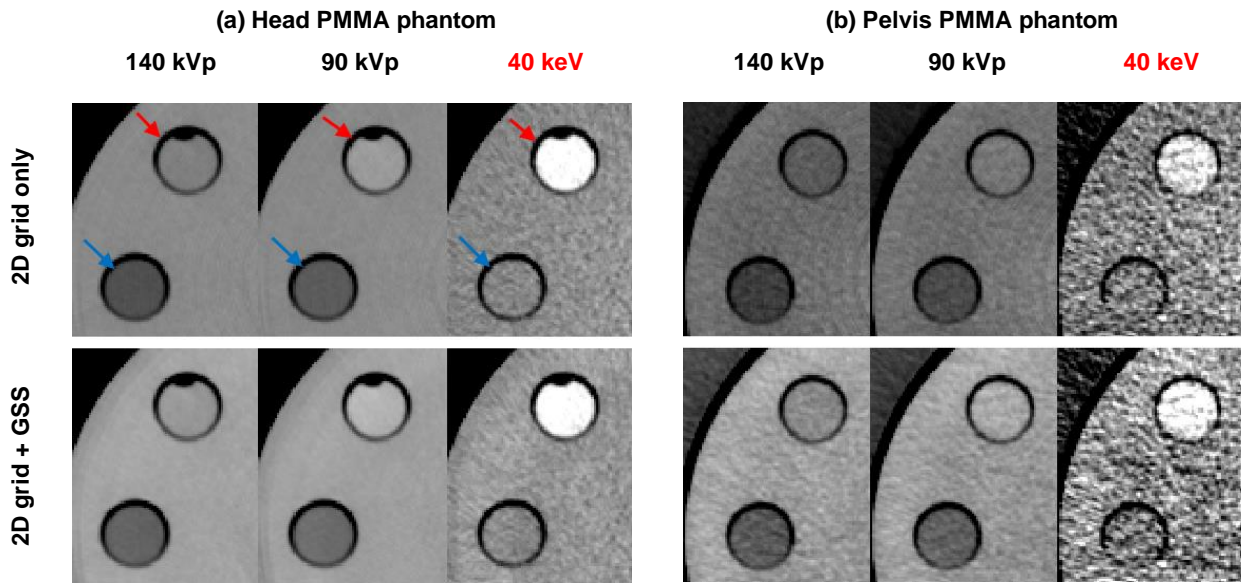

Fig. 10. Contrast visualization in polyenergetic and VME CBCT images. The ROI shows 2 and 5 mg/ml iodine vials (blue and red arrows, respectively). ROI locations in head (a) and pelvis (b) PMMA phantoms are shown in Fig. 8. HU window: [-100 300]

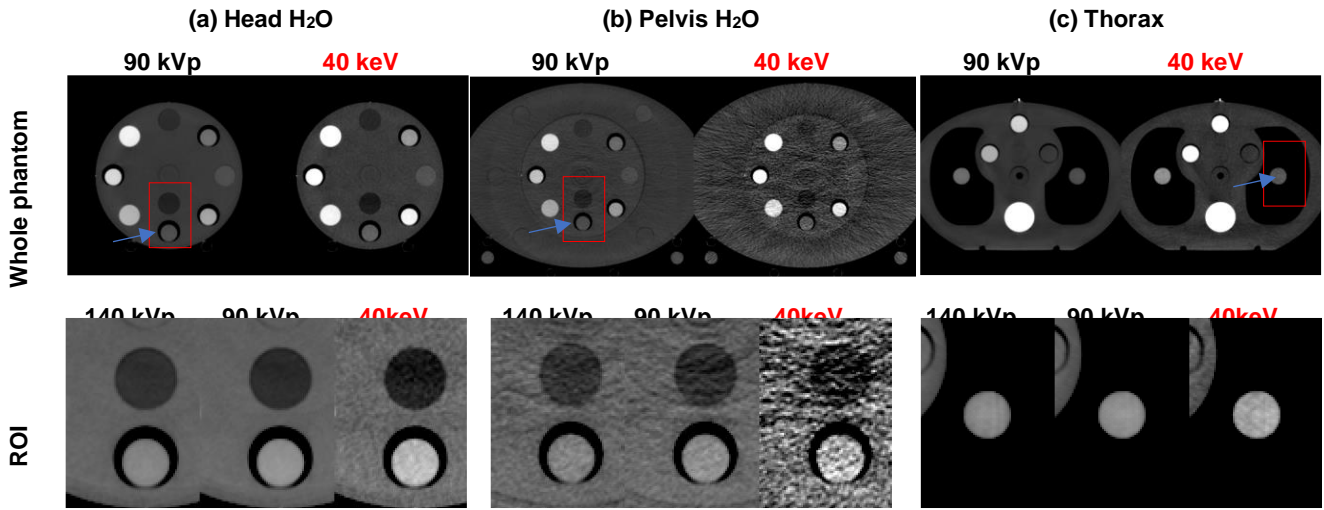

Fig. 11. Low contrast visualization in polyenergetic and VME CBCT images. ROIs containing 2 mg/ml iodine vial (blue arrow) are shown in the bottom row. All CBCT images were corrected for scatter using the GSS method. HU window of phantom images: [-200 700] HU window of ROI images: [-150 250].

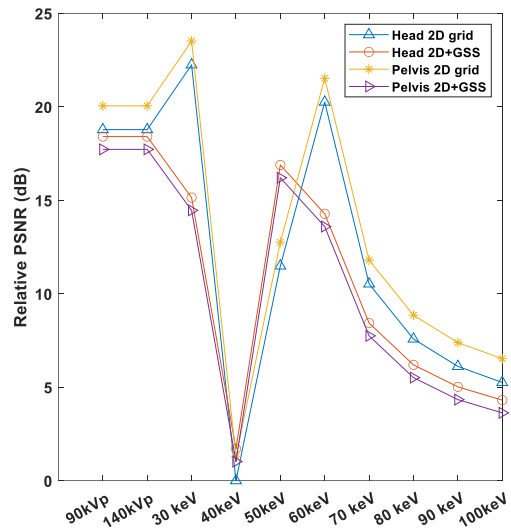

Fig. 12. Relative PSNR with respect to the minimum PSNR value in the head and pelvis H2O phantoms. The mean of 90 and 140 kVp images were used as the reference image in PSNR calculations.

## 4. Discussions

This study investigated the impact of 2D anti-scatter grids and GSS on DE material decomposition in CBCT images of human-torso-sized phantoms. The results indicate that the phantom size, and thereby the scatter fraction in projections, plays a critical role in the accuracy of DE decomposition. The suppression of scatter with a combination of a 2D anti-scatter grid and GSS scatter correction method can provide iodine quantification accuracy in DE CBCT images that is comparable to DE MDCT images.

While there are numerous other DE CBCT studies in the literature, material decomposition in human torso-sized objects has not been investigated previously. The vast majority of DE CBCT studies evaluated material quantification performance by using phantom diameters less than 20 cm,[4,11,14-17] where scatter fractions are lower. Such works demonstrated the measurement of iodine concentrations with less than 1 mg/ml iodine quantification errors. Iodine quantification errors were also less than 1 mg/ml in 20 cm diameter phantoms in this work, which agrees with the literature.

In the head and thorax-sized phantom projections, a 2D grid with a grid ratio of 12 can robustly reduce SPR below 0.2, and an iodine quantification accuracy of 1 mg/ml was achieved for iodine concentrations up to 40 mg/ml, which was comparable to the iodine quantification accuracy achieved in DE MDCT images. On the other hand, the SPR in pelvis phantom projections was a factor of 3 to 5 higher than for the head and thorax phantoms. When the 2D grid was the sole scatter suppression method, iodine quantification accuracy was acceptable at concentrations less than 10 mg/ml. Errors in iodine quantification accuracy reached 9.6 mg/ml when the iodine concentration was 40 mg/ml. This issue points to the fact that material decomposition accuracy highly depends on the scatter fraction in the projections.

The correction of scatter in conjunction with the 2D grid provided a substantial improvement in material decomposition accuracy in pelvis phantoms. Among the pelvis phantoms and all iodine concentrations evaluated, iodine quantification errors due to the change in phantom size were less than 1.5 mg/ml for the 2D grid+GSS approach, which was comparable to the error of 0.9 mg/ml in DE MDCT images. Thus, robust scatter mitigation is essential for achieving accurate DE CBCT material decomposition in the pelvis or abdomen, as demonstrated in DE CBCT scans acquired with a 2D grid and corrected with the GSS method.

The material filtering in low/high kVp HU correlation plots helped to differentiate materials with similar HU values. This filtering approach worked better in 2D-grid CBCT images corrected with the GSS method. In particular, the 2D grid+GSS approach improved the separation of bone-like objects and iodine in the HU correlation plot and helped to suppress some of the bone inserts in iodine images. However, not all bone voxels were filtered out from iodine images. This problem was partly due to higher noise and inherently close attenuation coefficients of bone and iodine in the 90 and 140 kVp CBCT scans. This problem can potentially be mitigated by reducing the spectral overlap between low and high kVp scans and implementing noise reduction methods.

Although scatter correction with the GSS method was crucial for achieving accurate iodine quantification, it led to noise amplification. This problem was further amplified in basis material images after DE processing. Moreover, noise amplification was a significant issue in VME images, which prevented any improvement in CNR when compared to polyenergetic 90 kVp images. Despite the lack of improvement in CNR, VME images showed substantially improved visualization of iodine contrast in qualitative visual comparisons with polyenergetic images. These observations are consistent with prior studies on iodine CNR in VME images.[3,16] Although the CNR values were comparable between VME and polyenergetic images, iodine visualization was superior in VME images. This observation suggests that the use of more realistic observer models, instead of relying solely on CNR, may be necessary to quantify low contrast visualization in VME images accurately.

The VME image quality trends observed in this study were consistent with previous research on VME CBCT, which mainly used phantom diameters of 20 cm or less.[11,15,16,40] It was found that for energies above the K edge of iodine, CNR and PSNR were maximized at 50 to 60 keV regardless of the phantom size, which is consistent with the literature. Furthermore, the improvement of CNR in VME images without additional noise suppression was minimal, or even less, compared to polyenergetic images. Previous studies have employed noise reduction and iterative reconstruction methods to overcome this issue.[11,16] Although only the FDK method was utilized in this study, iterative reconstruction methods could be explored in the future to improve the low-contrast visualization performance in VME imaging.

In this work, sequential low and high kVp scanning was employed because the available CBCT system did not have the capability of fast kV switching or dual-layer detector options. Although sequential scanning is not suitable for the clinical implementation of DE CBCT, it was sufficient for studying the effects of scatter and robust scatter suppression methods in the context of DE CBCT using static phantoms.

Another image quality challenge with 2D grid-only images is the ring artifacts. They are particularly visible in 140 kVp images of the pelvis $H_2O$ phantom (Fig. 5). Such ring artifacts are primarily caused by the residual scatter. When residual scatter is present, gain map correction does not fully suppress the wall shadows of the 2D grid due to the multiplicative nature of gain map correction and the additive nature of the residual scatter signal. After correction of residual scatter with the GSS method, the gain map correction step can more effectively suppress wall shadows, and hence, ring artifacts were reduced in CBCT images.

The use of sequential single-energy CBCT scans in this study was meant to simulate DE CBCT scans, but it is expected that the 2D grid will perform well in more realistic DE acquisition techniques, such as fast kVp switching and multilayer detectors. This is because the scatter rejection capability of the 2D grid is not likely to be affected by kVp switching or the layers of a multilayer detector. However, the performance of the GSS method in correcting inter-layer scatter in a multilayer detector or correcting scatter during rapid kVp switching remains to be investigated.

Several improvements can be implemented to improve DE CBCT imaging with a 2D grid and GSS scatter correction approach. First, energy separation in 90 and 140 kVp CBCT scans was smaller than the energy separation in MDCT scans. The energy separation between high and low kVp CBCT spectra can be enhanced by utilizing beam filters and removing lower energy X-rays in the 140 kVp beam. Second, only water-equivalent beam hardening was employed in CBCT scans. Material-specific beam hardening correction can further reduce beam hardening artifacts and improve basis material decomposition. Third, noise amplification in CBCT images after scatter correction is a barrier to achieving better DE processing performance. Noise reduction methods, such as iterative reconstruction techniques, can potentially improve DE imaging performance. Optimization of dose allocation for low/high kV scans can also help to manage noise characteristics in VME images. In this work, dose allocation was not investigated due to detector saturation in the linac-mounted CBCT system. Fourth, residual ring artifacts might be present due to nonlinear detector pixel response at low imaging doses[41] and incomplete correction of the grid wall shadows. These artifacts particularly affect the image quality in the central region of CBCT axial slices. To assess the clinical impact of the proposed methods, higher imaging doses and a ring artifact suppression method should be used. Lastly, image domain DE processing was employed in this work. This was partly due to reference MDCT scans used in a subset of investigations. Since projection data was unavailable for MDCT scans, image domain DE processing was selected for both MDCT and CBCT scans. Projection domain DE processing can potentially help reduce beam hardening artifacts in DE CBCT images.

## 5. Conclusions

In the past decade, DE CBCT has been actively investigated for applications in diagnostic imaging,[7] interventional imaging,[15,16] and image-guided radiation therapy.[8,11,24,42] However, the feasibility of DE CBCT material decomposition in the human torso has not yet been demonstrated in the literature, mainly due to the high intensity of scattered radiation.

The findings of this study suggest that DE CBCT material decomposition in the abdomen and pelvis may be feasible by using 2D anti-scatter grids and GSS to mitigate scatter. The proposed approach successfully decomposed high-scatter fluence images into iodine and water basis materials, with comparable iodine quantification accuracy to DE MDCT. Additionally, VME images of pelvis phantoms showed significant improvement in iodine contrast visualization. Overall, these results suggest that DE CBCT may have the potential to provide accurate and high-quality imaging in the abdomen and pelvis.

In CBCT-guided radiation therapy, the proposed DE approach with robust scatter mitigation has the potential to find numerous applications, such as improving the visualization of low contrast tissues during target localization and enabling more accurate calculation of electron density and proton stopping power ratios for CBCT-based dose calculations,[42] reduction of metal artifacts,[8] contrast agent enhanced tumor localization,[12] and improved extraction of radiomics features to predict tumor response and toxicity.[42]

**Acknowledgements**

This work was funded in part by grants from NIH/NCI R21CA198462 and R01CA245270.

**Conflict of Interest Statement**

The author does not have relevant conflicts of interest to disclose.

**References**

1. McCollough CH, Leng S, Yu L, Fletcher JG. Dual-and multi-energy CT: principles, technical approaches, and clinical applications. *Radiology.* 2015;276(3):637-653.

2. Matsumoto K, Jinzaki M, Tanami Y, Ueno A, Yamada M, Kuribayashi S. Virtual monochromatic spectral imaging with fast kilovoltage switching: improved image quality as compared with that obtained with conventional 120-kVp CT. *Radiology.* 2011;259(1):257-262.

3. Yu L, Leng S, McCollough CH. Dual-energy CT–based monochromatic imaging. *American journal of Roentgenology.* 2012;199(5_supplement):S9-S15.

4. Zbijewski W, Gang GJ, Xu J, et al. Dual-energy cone-beam CT with a flat-panel detector: Effect of reconstruction algorithm on material classification. *Medical Physics.* 2014;41(2):-.

5. Ding H, Ducote JL, Molloi S. Measurement of breast tissue composition with dual energy cone-beam computed tomography: A postmortem study. *Medical Physics.* 2013;40(6Part1):061902.

6. Tacher V, Radaelli A, Lin M, Geschwind J-F. How I do it: cone-beam CT during transarterial chemoembolization for liver cancer. *Radiology.* 2015;274(2):320-334.

7. Zbijewski W, Sisniega A, Stayman J, et al. Dual-energy imaging of bone marrow edema on a dedicated multi-source cone-beam CT system for the extremities. Paper presented at: Medical Imaging 2015: Physics of Medical Imaging2015.

8. Skaarup M, Edmund JM, Dorn S, Kachelriess M, Vogelius IR. Dual-energy material decomposition for cone-beam computed tomography in image-guided radiotherapy. *Acta Oncologica.* 2019;58(10):1483-1488.

9. Landry G, Reniers B, Granton PV, et al. Extracting atomic numbers and electron densities from a dual source dual energy CT scanner: experiments and a simulation model. *Radiotherapy and Oncology.* 2011;100(3):375-379.

10. Kruis MF. Improving radiation physics, tumor visualisation, and treatment quantification in radiotherapy with spectral or dual-energy CT. *Journal of applied clinical medical physics.* 2022;23(1):e13468.

11. Cassetta R, Lehmann M, Haytmyradov M, et al. Fast-switching dual energy cone beam computed tomography using the on-board imager of a commercial linear accelerator. *Physics in Medicine & Biology.* 2020;65(1):015013.

12. Jones BL, Altunbas C, Kavanagh B, Schefter T, Miften M. Optimized dynamic contrast-enhanced cone-beam CT for target visualization during liver SBRT. Paper presented at: Journal of Physics: Conference Series2014.

13. Müller K, Datta S, Ahmad M, et al. Interventional dual-energy imaging—Feasibility of rapid kV-switching on a C-arm CT system. *Medical physics.* 2016;43(10):5537-5546.

14. Jiang X, Fang C, Hu P, Cui H, Zhu L, Yang Y. Fast and effective single-scan dual-energy cone-beam CT reconstruction and decomposition denoising based on dual-energy vectorization. *Medical physics.* 2021;48(9):4843-4856.

15. Shi L, Lu M, Bennett NR, et al. Characterization and potential applications of a dual-layer flat-panel detector. *Medical physics.* 2020;47(8):3332-3343.

16. Ståhl F, Schäfer D, Omar A, et al. Performance characterization of a prototype dual-layer cone-beam computed tomography system. *Medical Physics.* 2021;48(11):6740-6754.

17. Wang W, Ma Y, Tivnan M, et al. High-resolution model-based material decomposition in dual-layer flat-panel CBCT. *Medical physics.* 2021;48(10):6375-6387.

18. Müller K, Ahmad M, Spahn M, et al. Towards material decomposition on large field-of-view flat panel photon-counting detectors—first in-vivo results. Paper presented at: Proceedings of The Fourth International Conference on Image Formation in X-Ray Computed Tomography2016.

19. Ahmad M, Fahrig R, Pung L, et al. Assessment of a photon-counting detector for a dual-energy C-arm angiographic system. *Medical physics.* 2017;44(11):5938-5948.

20. Ji X, Feng M, Treb K, Zhang R, Schafer S, Li K. Development of an integrated C-arm interventional imaging system with a strip photon counting detector and a flat panel detector. *IEEE transactions on medical imaging.* 2021;40(12):3674-3685.

21. Schmidt TG. CT energy weighting in the presence of scatter and limited energy resolution. *Medical physics.* 2010;37(3):1056-1067.

22. Zhao C, Liu SZ, Wang W, et al. Effects of x-ray scatter in quantitative dual-energy imaging using dual-layer flat panel detectors. Paper presented at: Medical Imaging 2021: Physics of Medical Imaging2021.

23. Zhang T, Chen Z, Zhou H, Bennett NR, Wang AS, Gao H. An analysis of scatter characteristics in x-ray CT spectral correction. *Physics in Medicine & Biology.* 2021;66(7):075003.

24. Schröder L, Stankovic U, Rit S, Sonke J-J. Image quality of dual-energy cone-beam CT with total nuclear variation regularization. *Biomedical Physics & Engineering Express.* 2022;8(2):025012.

25. Jiang X, Cui H, Liu Z, Zhu L. Residual W-shape network (ResWnet) for dual-energy cone-beam CT imaging. Paper presented at: 7th International Conference on Image Formation in X-Ray Computed Tomography2022.

26. Zhu J, Su T, Zhang X, et al. Feasibility study of three-material decomposition in dual-energy cone-beam CT imaging with deep learning. *Physics in Medicine & Biology.* 2022;67(14):145012.

27. Wang AS. Single-shot quantitative x-ray imaging from simultaneous scatter and dual energy measurements: a simulation study. Paper presented at: Medical Imaging 2021: Physics of Medical Imaging2021.

28. Gao H, Zhang T, Bennett NR, Wang AS. Densely sampled spectral modulation for x-ray CT using a stationary modulator with flying focal spot: a conceptual and feasibility study of scatter and spectral correction. *Medical Physics.* 2021;48(4):1557-1570.

29. Pivot O, Fournier C, Tabary J, Létang JM, Rit S. Scatter correction for spectral CT using a primary modulator mask. *IEEE transactions on medical imaging.* 2020;39(6):2267-2276.

30. Altunbas C, Kavanagh B, Alexeev T, Miften M. Transmission characteristics of a two dimensional antiscatter grid prototype for CBCT. *Medical physics.* 2017;44(8):3952-3964.

31. Alexeev T, Kavanagh B, Miften M, Altunbas C. Two-dimensional antiscatter grid: A novel scatter rejection device for Cone-beam computed tomography. *Medical physics.* 2018;45(2):529-534.

32. Yu Z, Park Y, Altunbas C. Simultaneous scatter rejection and correction method using 2D antiscatter grids for CBCT. Paper presented at: Medical Imaging 2020: Physics of Medical Imaging2020.

33. Altunbas C, Park Y, Yu Z, Gopal A. A Unified Scatter Rejection and Correction Method for Cone Beam Computed Tomography. *Medical Physics.* 2020. doi: 10.1002/mp.14681.

34. Bayat F, Eldib ME, Kavanagh B, Miften M, Altunbas C. Concurrent kilovoltage CBCT imaging and megavoltage beam delivery: suppression of cross-scatter with 2D antiscatter grids and grid-based scatter sampling. *Physics in Medicine & Biology.* 2022;67(16):165005.

35. Altunbas C, Alexeev T, Miften M, Kavanagh B. Effect of grid geometry on the transmission properties of 2D grids for flat detectors in CBCT. *Physics in Medicine & Biology.* 2019;64(22):225006.

36. Park Y, Alexeev T, Miller B, Miften M, Altunbas C. Evaluation of scatter rejection and correction performance of 2D antiscatter grids in cone beam computed tomography. *Medical Physics.* 2021;48(4):1846-1858.

37. Roos PG, Colbeth RE, Mollov I, et al. Multiple-gain-ranging readout method to extend the dynamic range of amorphous silicon flat-panel imagers. *Proc SPIE.* 2004;5368:139-149.

38. Biguri A, Dosanjh M, Hancock S, Soleimani M. TIGRE: a MATLAB-GPU toolbox for CBCT image reconstruction. *Biomedical Physics & Engineering Express.* 2016;2(5):055010.

39. Wang G. X-ray micro-CT with a displaced detector array. *Medical physics.* 2002;29(7):1634-1636.

40. Shi L, Bennett NR, Shapiro E, et al. Comparative study of dual energy cone-beam CT using a dual-layer detector and kVp switching for material decomposition. Paper presented at: Medical Imaging 2020: Physics of Medical Imaging2020.

41. Altunbas C, Lai C-J, Zhong Y, Shaw CC. Reduction of ring artifacts in CBCT: Detection and correction of pixel gain variations in flat panel detectors. *Medical physics.* 2014;41(9):091913.

42. Sajja S, Lee Y, Eriksson M, et al. Technical principles of dual-energy cone beam computed tomography and clinical applications for radiation therapy. *Advances in radiation oncology.* 2020;5(1):1-16.